\documentclass[11pt,a4paper,english,superscriptaddress,prd,aps,preprint]{revtex4}
\usepackage{amssymb}
\usepackage{amsmath} 
\usepackage{slashed}
\usepackage{graphicx}
\usepackage{babel}
\usepackage{hyperref}
\usepackage{color}
\makeatletter
\AtBeginDocument{\let\@elt\relax}\makeatother

\begin{document}

\title{Diamonds in Klein geometry}

\author{{Rafael Mancini Santos \footnote{Currently at Divisão de Astrofísica, Instituto Nacional de Pesquisas Espaciais - Inpe, São Paulo, Brazil} }}\email{rafael.mancini@inpe.br }

\author{L. C. T. Brito}
\email{lcbrito@ufla.br}
	
\author{{Cleverson Filgueiras} }\email{cleverson.filgueiras@ufla.br}

\affiliation{Departamento de Física, Universidade Federal de Lavras, Campus Universitário, Lavras,3037, Minas Gerais, Brasil}

\begin{abstract}
Recently, there has been a suggestion that the Unruh effect might manifest in metamaterials at accessible Unruh temperatures. In certain instances, the class of metamaterials that could be instrumental for this observation exhibits a Klein signature instead of a Minkowski one. Consequently, confirming this effect in those materials necessitates a more meticulous analysis.
In this paper, we employ the path integral formulation of Quantum Field Theory to investigate the analogue of the Unruh effect in Kleinian geometry. We perform calculations for a scalar theory, provided we restrict the action to a convenient subspace of the Kleinian spacetime. As a result, we determine the diamond temperature for a static observer with a finite lifetime. The outcome suggests that metamaterials could serve as a potential system for observing diamond regions.
\end{abstract}

\maketitle

\section{Introduction}

The exchange of ideas between Quantum Field Theory (QFT) and condensed matter physics has been very fruitful over the years. The application of the concept of spontaneously broken symmetry in particle physics \cite{Nambu:1960xd} and superconductivity \cite{Anderson:1963pc} in the early 1960s represents the first remarkable example. A more recent case is the discovery that two-dimensional graphene is described at low energy by the massless Dirac equation \cite{graphene}. Besides these examples, there exists a wide range of ideas connecting particle physics and condensed matter \cite{Wilczek:2016jqq}. In this paper, we will investigate analogous phenomena in QFT and condensed matter physics motivated by a recent proposal that suggests the Unruh effect might occur in certain classes of metamaterials \cite{Smolyaninov:2020noj}.

The Unruh effect is a prediction of QFT \cite{Unruh:1976db}, and therefore it can be interpreted as a natural consequence of Quantum Mechanics and Special Relativity. It states that uniformly accelerating observers experience the Minkowski vacuum as a thermal bath with temperature $T = \hbar a/2\pi c \kappa$, where $a$ is the observer's acceleration,  $ \hbar$ is the reduced Planck constant,  $c$ is the speed of light, and $\kappa$ is the Boltzmann's constant. A detailed review on the fundamentals and applications of the Unruh effect can be found in reference \cite{Crispino:2007eb}. 

Direct experimental verification of the Unruh effect is a difficult challenge, as an acceleration of order $10^{20}\,\text{m/s}^2$ is required to achieve a temperature of the order of $1 {\text K}$.  Nevertheless, there are good reasons to believe that it is possible to detect the Unruh effect using both quantum \cite{Martin-Martinez:2010gnz,martin2013berry} and classical \cite{PhysRevLett.118.161102} systems, with experiments that may be accessible to the current technology. Also, the reported observation of thermalization at the Unruh temperature caused by radiation emitted from accelerated positron might be the first direct measure of the Unruh temperature in a high-energy system \cite{PhysRevD.104.025015}. 

Recently it was suggested that the Unruh effect might occur in hyperbolic metamaterial waveguides \cite{Smolyaninov:2020noj}. These materials have a high degree of anisotropy, and photons propagating through them might achieve high accelerations, which could be sufficient to put the Unruh temperature at values accessible to experiments. However, it is possible to show that hyperbolic metamaterials may behave effectively as a space with a Kleinian signature as light propagates through it \cite{smolyaninov2010metric}. In this case, a careful analysis of the existence of the Unruh effect is necessary.

Investigation concerning the Kleinian signature goes
back to \cite{alty1994kleinian}. Physical effects in optics and cosmology in the context of analog systems were discussed in  \cite{fumeron2015optics,figueiredo2016modeling}. It was only in reference \cite{PhysRevD.103.044023} that the Kleinian geometry was investigated in more detail. In particular, the authors show that acceleration restricts the time coordinate of a classical particle to take values within an interval determined by the magnitude of acceleration. As we will see, this restriction is crucial for obtaining the analogous Unruh temperature in Klein geometry. Besides, only phenomena in relativistic quantum mechanics were considered in \cite{PhysRevD.103.044023}.

A traditional way to deduce the Unruh effect for free scalar fields in Minkowski space is by using the canonical formalism of QFT ( see \cite{Crispino:2007eb} for a pedagogical and \cite{wald1} for a more rigorous presentation).
In particular, Minkowski space is globally hyperbolic and static (see \cite{wald2} for details about classical solutions of the Klein-Gordon equation in hyperbolic spacetimes). These characteristics justify the canonical formulation of QFT for free scalar fields in Minkowski spacetime. As a consequence, it is possible to expand the quantum field in terms of creation and annihilation operators and construct the Hilbert space as a Fock space with a well-defined vacuum state that preserves translational symmetry. The relevant point for our discussion is that Klein's geometry, in which we are interested, is ultrahyperbolic instead of hyperbolic \cite{Tegmark:1997jg} .  

In this paper, we utilize the method of path integrals of QFT to investigate and provide justification for the occurrence of an effect analogous to the Unruh effect in an effective geometry with a Kleinian signature. In particular, the method circumvents the mathematical challenge of constructing a Hilbert space for QFT in ultrahyperbolic geometries. We calculate the counterpart to the Unruh temperature in this scenario, the diamond temperature, and discuss why metamaterials are a potential candidate for providing insights into the presence of diamond regions \cite{Martinetti:2002sz,Chakraborty:2022qdr}. Given the recent observation of photon acceleration in tapered optical waveguides \cite{ge2021observation}, we may have the possibility of testing the occurrence of the Unruh or analogous effects in a setup similar to the one outlined in reference \cite{Smolyaninov:2020noj}.

The paper is organized as follows. In section II, we review the deduction of the Unruh effect by using the path integral approach of QFT \cite{unruh1984acceleration} and show how to obtain the diamond temperature with this approach in the Minkowski space. In section III, we apply path integrals to a geometry with a Kleinian signature to obtain the  diamond temperature. In section IV we discuss the connection with metamaterials. Finally, in section V, we summarize our results and make some final remarks. 

Through the rest of the paper we will use units in which $c = \hbar = \kappa = 1$.

\section{Path integrals,  Unruh effect and the diamond temperature}

We begin our discussion with the generating functional for the Green function in the path integral formulation of QFT: 
\begin{equation}\label{generatingZ}
Z[J]=\rho\int D\phi e^{i\left(S+\int d^{4}xJ\phi\right)},
\end{equation}
where the normalization constant $\rho$ is chosen such that  $Z[0]=1$. We are interested in the theory of a scalar field $\phi$ whose dynamic is governed by the classical action 
\begin{equation}\label{action}
S=\int_{t_1}^{t_2}dt\int_{V} d^{3}x \mathcal{L},    
\end{equation}
with the Lagrangian density
\begin{equation}
\mathcal{L}=-\frac{1}{2}\left(g_{\mu\nu}\partial^{\mu}\phi\partial^{\nu}\phi+m^{2}\phi^{2}\right) \label{densityL}.
\end{equation}
Here, $m^{2}$ is a parameter with the dimension of mass squared, and $g_{\mu\nu}$ is the metric for a flat geometry. In the present section, we are interested in the Unruh effect in the Minkowski space. So we consider $g_{\mu\nu} = (-1, 1, 1, 1)$, leading to the Lagrangian density
\begin{equation}\label{KG}
\mathcal{L}=\frac{1}{2}\left(\frac{\partial\phi}{\partial t}\right)^{2}-\left(\nabla\phi\right)^{2}-m^{2}\phi^{2}.
\end{equation}

Also, we need the classical Lagrangian density as defined by a non-inertial observer with constant acceleration $a$. It is obtained by the Rindler transformation of coordinates
\begin{equation}\label{rindlerT}
x=r\cosh\left(a\eta\right)\,\,\,\,\,\,\,\,\,\,{\rm and} \,\,\,\,\,\,\,\,\,\, t=r\sinh\left(a\eta\right), \end{equation}
where $0 < r < \infty$ and $-\infty < \eta < +\infty$ are the coordinates correspondent to the geometry of the right wedge of the Rindler space:
\begin{equation}
d^{2}s = -a^{2} r^{2} d\eta^2 + dr^2 + dy^2 + dz^2. 
\end{equation}
Thus, an observer with acceleration $a$ and fixed spacial coordinates $r, y$, and $z$ measures the proper time $d\eta$ at $r = \frac{1}{a}$. Using the transformation (\ref{rindlerT}) in the Lagrangian density  (\ref{densityL}), we obtain the Lagrangian density for such an observer:
\begin{equation}\label{lagrangianR}
 \mathcal{L}^{R} =\frac{1}{2}\,a\,r\,\left[\frac{1}{\left(ar\right)^{2}}\left(\frac{\partial\phi}{\partial\eta}\right)^{2}-\left(\frac{\partial\phi}{\partial r}\right)^{2}-\left(\nabla_{\perp}\phi\right)^{2}-m^{2}\phi^{2}\right].   
\end{equation}
Here, we have $\nabla_{\perp} = (\frac{\partial}{\partial y},\frac{\partial}{\partial z})$.

Now, we turn our attention to quantum theory. We will consider the QFT in terms of an imaginary time. By taking $t_1 = -i\tau$ and $t_2 = i\tau$  and letting  $\tau \rightarrow \infty$, the generating functional (\ref{generatingZ}) corresponds to the standard QFT at zero temperature. On the other hand, to introduce the equilibrium temperature   $T$ into the QFT defined by (\ref{generatingZ}), we must consider the time coordinate $t$  as an imaginary number  and confine it to the interval 
\begin{equation}\label{imag_time}
t_2 - t_1 = i\beta, 
\end{equation}
with $\beta = \frac{1}{T}$. 
 
It is a standard procedure performing   a change of coordinates $t \rightarrow -i\tau$ in (\ref{generatingZ}) such that the generating functional in the QFT at finite temperature $T$ can be written as
\begin{equation}\label{functionalTau}
Z[J]=\rho\int D\phi e^{-\int_{-\beta/2}^{\beta/2}d\tau\int d^{3}x\left(\mathcal{L}_{\tau}+J\phi\right)}.
\end{equation}
Here, we have 
\begin{equation}\label{prescrip}
\mathcal{L}_{\tau} = - \mathcal{L}(t=-i\tau),    
\end{equation} 
which means that on the right-hand side of this expression we need to replace $t\rightarrow -i\tau$ in the Lagrangian density (\ref{densityL}). Convergence of the functional generator (\ref{functionalTau}) is assured in Minkowski space when utilizing the Euclidean action. For geometries where convergence does not occur for all field configurations, like the Klein geometry we will consider in the next section, the sum in the path integrals must take only the configurations for which the functional generator is convergent.

Let us now apply the prescription outlined above to review the calculation of the Unruh temperature in the  Minkowski space \cite{unruh1984acceleration}. Utilizing the prescription as indicated in (\ref{prescrip}), we can derive the Lagrangian densities for the inertial and  non-inertial observers from (\ref{KG}) and (\ref{lagrangianR}), which are given respectively by 
\begin{equation}\label{inertialL}
\mathcal{L}_{\tau}=\frac{1}{2}\left[\left(\frac{\partial\phi}{\partial\tau}\right)^{2}+\left(\nabla\phi\right)^{2}+m^{2}\phi^{2}\right].
\end{equation}
and
\begin{equation}\label{RindlerTau}
\mathcal{L}_{\tau}^{R}=\frac{1}{2}\left[\frac{1}{\left(ar\right)^{2}}\left(\frac{\partial\phi}{\partial\tau}\right)^{2}+\left(\frac{\partial\phi}{\partial r}\right)^{2}+\left(\nabla_{\perp}\phi\right)^{2}+m^{2}\phi^{2}\right].
\end{equation}

The QFT in  Rindler spacetime at temperature $T$ is defined by the generating functional (\ref{functionalTau}) with the action
\begin{equation}\label{actionTau}
S_{\tau}^{R}  =  \frac{1}{2} \int_{0}^{\infty}dr\int_{-\beta/2}^{+\beta/2}d\tau\int_{x,y} \,a\,r\,\left[\frac{1}{\left(ar\right)^{2}}\left(\frac{\partial\phi}{\partial\tau}\right)^{2}+\left(\frac{\partial\phi}{\partial r}\right)^{2}+\left(\nabla_{\perp}\phi\right)^{2}+m^{2}\phi^{2}\right],
\end{equation}
where the Lagrangian (\ref{RindlerTau}) has been employed. Here, we have introduced the symbol $\int_{y,z}$ to denote integrals that do not change by the transformation (\ref{rindlerT}). Changing variables in (\ref{actionTau})  as
\begin{equation}\label{polar}
x=r\cos (a\tau)\,\,\,\,\,\,{\rm and} \,\,\,\,\,\, t=r\sin (a\tau),
\end{equation}
with $-\beta/2 < \tau < +\beta/2$, we find that if we take  
\begin{equation}\label{beta}
\beta = \frac{2\pi}{a},
\end{equation}
we will have an integral in polar coordinates over the entire plane.

Actually, using (\ref{polar}) and (\ref{beta}) in (\ref{actionTau}), we obtain
\begin{equation}
S_{\tau}^{R}=\frac{1}{2}\int_{-\infty}^{+\infty}dt\int_{-\infty}^{+\infty}dx\int_{y,z} \left[\left(\frac{\partial\phi}{\partial t}\right)^{2}+\left(\frac{\partial\phi}{\partial x}\right)^{2}+\left(\nabla_{\perp}\phi\right)^{2}+m^{2}\phi^{2}\right].
\end{equation}
Remarkably, this is the action $S_{\tau} = \int d^{4}x\mathcal{L}_{\tau}$ that defines the generating function for a QFT at zero temperature with the  Lagrangian density (\ref{inertialL}), for an inertial observer. Hence, the QFT at zero temperature for an inertial observer is equivalent to the QFT at temperature $T = \frac{a}{2\pi}$ for a Rindler observer.

Interestingly, the method can be used to calculate the temperature when the QFT at zero temperature is restricted to a finite region of spacetime. In particular, if we restrict the spacetime in (\ref{action}) to be the right wedge of the light cone, the angle $a\tau$ in (\ref{polar}) must take values in the interval $-\frac{\pi}{4a} < \tau < \frac{\pi}{4a}$. So, instead of (\ref{beta}), the choice of $\beta$ in this case is given by
\begin{equation}\label{beta_diamond}
\beta_{D} = \frac{\pi}{2a},
\end{equation}
which is equivalent to the temperature, $T_D = \frac{2 a}{\pi}$. As we will discuss in section 4, this is the temperature experienced by a static observer with a finite lifetime $\mathcal{T} = \frac{1}{a}$ in a diamond region \cite{Martinetti:2002sz}.  

\section{The diamond temperature in Klein geometry}
In the previous section, we outlined a general methodology to obtain the  Unruh temperature for a Rindler observer and for a observer with a finite lifetime in Minkowski spacetime. Now, we will use this methodology to focus on our primary objective of determining the analogous result for Klein Geometry. 

The Klein geometry corresponds to the metric $g_{\mu\nu} = (-1, -1, 1, 1)$, with the line element
\begin{equation}\label{kleinian_metric}
d^{2}s = - dt^{2}- dx^{2} + dy^{2} + dz^{2}.    
\end{equation}
Using this metric in the Lagrangian density (\ref{densityL}), we have the theory of a scalar field in the Klein space:
\begin{equation}\label{KleinScalar}
\mathcal{L}^{K}=\frac{1}{2}\left[\left(\frac{\partial\phi}{\partial t}\right)^{2}+ \left(\frac{\partial\phi}{\partial x}\right)^{2}-\left(\nabla_{\perp}\phi\right)^{2}-m^{2}\phi^{2}\right].
\end{equation}

The coordinate transformation to a system with constant acceleration $a$ in Klein space can be described by
\begin{equation}\label{TransfKlein}
x=r\cos a\eta\,\,\,\,\,\,{\rm and}\,\,\,\,\,\, t=r\sin a\eta,
\end{equation}
where $0 < r < \infty$ and $-\frac{\pi}{4} < a\eta < +\frac{\pi}{4}$. Notably, the first relevant point for the discussion of the Unruh effect in the case of Klein space is that the ``time" coordinate $\eta$ of the accelerated observer is initially confined. Furthermore, the coordinates do not cover the full plane but only the right wedge of spacetime. As we will see, the coordinate $\eta$ is defined in the interval $(-\frac{\pi}{4a},\frac{\pi}{4a})$ to ensure equivalence between the QFT at zero temperature and the QFT at the finite temperature. The restriction of time in the motion of an accelerated particle was previously obtained in a classical context in \cite{PhysRevD.103.044023}. In the context of the equivalence between zero and finite-temperature QFTs we consider here, it is crucial to impose a restricted interval of time. The necessity to take this restriction as coinciding with the right wedge of spacetime will become clear soon. 

The Lagrangian density for the accelerated observer in Klein space is given by:
\begin{equation}\label{LagrangAKein}
\mathcal{L}^{(a)} =\frac{1}{2}\,a\,r\,\Bigg[\frac{1}{(ar)^{2}}\left(\frac{\partial\phi}{\partial \eta}\right)^{2}+\left(\frac{\partial\phi}{\partial r}\right)^{2}-(\nabla_{\perp}\phi)^{2}\Bigg],    
\end{equation}
which is obtained by using the coordinates transformation (\ref{TransfKlein}) in
(\ref{KleinScalar}).

In the QFT defined by the generating functional (\ref{functionalTau}), we require the Lagrangian densities  (\ref{KleinScalar}) and (\ref{LagrangAKein}) in the imaginary time. These Lagrangian densities  are given respectively by
\begin{equation}\label{inertialKleinIm}
 \mathcal{L}^{(K)}_{\tau} = \frac{1}{2} \left[\left(\frac{\partial\phi}{\partial \tau}\right)^{2}-\left(\frac{\partial\phi}{\partial x}\right)^{2}+(\nabla_{\perp}\phi)^{2}\right]   
\end{equation}
and
\begin{equation}
\mathcal{L}^{(a)}_{\tau} = \frac{1}{2}\,a\,r\,\Bigg[\frac{1}{(ar)^{2}}\left(\frac{\partial\phi}{\partial \tau}\right)^{2}-\left(\frac{\partial\phi}{\partial r}\right)^{2}+(\nabla_{\perp}\phi)^{2}\Bigg].
\end{equation}
By using (\ref{imag_time}) and the action for the accelerated frame,
\begin{equation}\label{action_accel_klein}
 S^{a}_{\tau} = \int_{-\frac{\pi}{4a}}^{+\frac{\pi}{4a}} d\tau \int_{0}^{\infty} dr \int_{y,z} \mathcal{L}^{(a)}_{\tau},   \end{equation}
 in the generating function (\ref{functionalTau}) we obtain  a QFT at the temperature $T_D = \frac{2 a}{\pi}$, which is the diamond temperature obtained from (\ref{beta_diamond}).
To demonstrate that this action is equivalent to the action defined by the Lagrangian density (\ref{inertialKleinIm}) of the "inertial" frame in Klein space, we need to make the following change of coordinates:
\begin{equation}\label{transform_klein}
x=r\cosh(a\tau)\,\,\,\,\,\,\,\,\,\,\,\, {\rm and}\,\,\,\,\,\,\,\,\,\,\,\,
t=r\coth\left(\frac{\pi}{4}\right)\sinh(a\tau).
\end{equation}
The factor $\coth\left(\frac{\pi}{4}\right)$ in the above expression for the $t$ coordinate enforce  $ x = \pm t $  at the limit $ \tau \rightarrow \pm \frac{\pi}{4a} $. Using the coordinate transformation (\ref{transform_klein})  in the action (\ref{action_accel_klein}), we obtain
\begin{equation}\label{action_zero_temp}
S^{(K)}_{\tau} = \int_{\mathcal{M}} d^4x \mathcal{L}^{(K)}_{\tau},
\end{equation}
where 
\begin{equation}\label{subspace}
 \mathcal{M}:=\{(t,x,y,z)\in\mathbb{R}^{2,2}\,;\,x>|t|\}   
\end{equation}
is the Klein space for the inertial theory, at zero temperature. The symbol $\mathbb{R}^{2,2}$ is the space defined from the Klenian metric (\ref{kleinian_metric}). As we said before, $\mathcal{M}$ covers only the right wedge of the spacetime. 

In conclusion, the equivalence between the zero-temperature QFT and the QFT at temperature $T_D = \frac{2a}{\pi}$ holds only when we restrict the space to the subspace defined in (\ref{subspace}). This restriction is necessary to establish an analogous of the Unruh temperature, the diamond temperature $T_D$, when the space has a Klein signature.

%%%%%%%%%%%%%%%%%%%%%%%%%%%%%%%%%%%%%%%%%%%%%%%%%
\section{Diamond regions and metamaterials}
%%%%%%%%%%%%%%%%%%%%%%%%%%%%%%%%%%%%%%%%%%%%%%%%%
Diamond is the name of a region encompassing the intersection of the lightcone of the future of the initial point and the lightcone of the past of the final point in  a finite  spacetime trajectory. In Minkowski spacetime, one can interpret the diamond as the region causally connected to a static observer with a finite lifetime $\mathcal{T}$ \cite{Martinetti:2002sz,Chakraborty:2022qdr}. As shown in the seminal paper \cite{Martinetti:2002sz}, an observer in this region sees a thermalization process. Remarkably, this thermalization does not vanish even for non-accelerating observers. In this case, the temperature is given by $T_D = \frac{2}{\pi \mathcal{T}}$, which is called the diamond temperature.

There is a general procedure mapping the diamond region in the Rindler wedge \cite{Chakraborty:2022qdr}, which is exactly the right wedge of the spacetime that we used before to establish the diamond temperature using the path integral approach of QFT. The interesting point we have shown is that a diamond temperature exists even when we consider a Klenian geometry instead of a Minkowski one. In particular, constraints in Klenian geometry must be necessarily imposed if we want to obtain an analogous for the Unruh temperature in this case.

The physical interpretation of the diamond temperature in Klein space follows the same interpretation in Minkowski spacetime. So, it is associated with some ``observer"  with a finite lifetime. The natural restrictions occurring in a geometry with Klein signature seem as if this geometry were the natural one to observe diamond regions. Actually, the map between the right wedge and the diamond region exists since the conformal symmetry can also be defined for the flat Klein space \cite{Atanasov:2021oyu}.

Interestingly, the fact that spacetimes with two-time coordinates have inherent instabilities has been discussed previously \cite{Tegmark:1997jg}.  The instabilities may be related, for example, to decay channels of some elementary particles that do not exist in a space with just one-time coordinate. Although there is no evidence about more than one time coordinate in spacetime, geometries like the Klein provides an interesting background to study processes where some kind of ``decay"  can be induced by geometry. Actually, as we can see from the metric signature introduced in section 3,  Klein space can be interpreted in the realm of physics with two-time coordinates. So, it is interesting to ask if some physical system could mimic the Klein geometry,  at least as an effective geometry. Anisotropic metamaterials seem like the more prominent physical system for this purpose.

Metamaterials are an exciting subject of research \cite{cui, Smolyaninov2017}. Interest in these materials is increasing due to the advent of technologies that make it possible to artificially engineer the structure of materials to obtain unusual properties, such as negative values of electric permittivity, which is described as tensor. This way, we find interesting physical possibilities. Specifically, anisotropic metamaterials can be engineered to possess a negative permittivity component, utilizing specific fabrication techniques\cite{PrintMeta,metaTheory,expe}. In this case, the permittivity tensor for such metamaterials can be expressed as:
\begin{equation} 
\varepsilon_{ij} = \begin{pmatrix} \varepsilon_1 & 0 & 0 \\ 0 & \varepsilon_1 & 0 \\ 0 & 0 & \varepsilon_2 \ \end{pmatrix}\;.
\end{equation}
Assuming $\epsilon_1 > 0$ and $\epsilon_2 < 0$, this tensor serves as an effective metric for electromagnetic waves. Consequently, an incident electromagnetic wave manifests an 'extraordinary' component that propagates akin to a real scalar field over a Kleinian spacetime. The dispersion relation in this context is given by:
\begin{equation} 
\frac{\omega^2}{c^2} = \frac{k_x^2}{\epsilon_1} + \frac{k_y^2}{\epsilon_1} - \frac{k_z^2}{|\epsilon_2|}\;,
\end{equation}
where ${\bf k} = (k_x, k_y, k_z)$ represents the wave vector, $\omega$ denotes the angular frequency, and $c$ is the speed of light.

In addition to optic metamaterials, other instances of anisotropic materials are explored. In a study focused on nematic liquid crystals \cite{fumeron2015optics}, the behavior of light rays is investigated within a distinct scenario. The effective metric in cylindrical coordinates is expressed as:
\begin{equation} 
ds_{\text{eff}}^2 = -c^2dt^2 + \varepsilon_0dr^2 + \varepsilon_er^2d\phi^2 + \varepsilon_0dz^2 \;. 
\end{equation}
This configuration assumes permittivities $\epsilon_0 > 0$ and $\epsilon_e < 0$, aligning with a Klein-like spacetime signature. Furthermore, it is noteworthy that Kleinian spacetime can also be simulated in the realm of electronic metamaterials\cite{figueiredo2016modeling}.

Besides, metamaterials have been used to obtain analogous effects arising from General Relativity \cite{Leonhardt:2006ai}, quantum gravity \cite{Smolyaninov:2022lio}, cosmology \cite{Smolyaninov:2011ah}, and to investigate modifications in Planck's law for thermal radiation \cite{BiehsPRL2015}. They also serve as a possible background that could enhance the Unruh effect \cite{Smolyaninov:2020noj}. This last proposal was that motivated the present investigation.

Actually, it has been suggested that Kleinian geometry can be established in certain types of metamaterials \cite{smolyaninov2010metric}. So, it is interesting to search for physical processes in these metamaterials that resemble some kind of "decay," which could be interpreted as a manifestation of a thermalization process occurring in a diamond region. Of course, it is not our intention to discuss this matter here. The point is that since metamaterials can mimic Klein geometry, and there is an analogous effect to the Unruh temperature in these geometries,  the diamond temperature, these materials naturally can be useful as a system for investigating diamond regions.

\section{Concluding Remarks}

In conclusion, this paper has explored the analogue of the Unruh effect in a Kleinian geometry, which is the "diamond temperature" associated with a finite spacetime trajectory. The path integral formulation of Quantum Field Theory (QFT) was employed to establish this temperature, which was found to be $T_D = \frac{2a}{\pi}$ for a static observer with a finite lifetime $\mathcal{T}= \frac{1}{a}$ in the diamond region. The study also highlighted the importance of restricting the spacetime geometry to a specific subspace for the existence of the diamond temperature.

Moreover, we suggest that hyperbolic metamaterials may serve as a suitable system to observe diamond regions, given their potential to mimic Kleinian geometry. This proposition opens up the possibility of experimental investigations into the concept of diamond temperatures in the context of metamaterials. As is known, metamaterials are artificial materials designed to possess properties not found in natural materials. Some types of metamaterials include negative-index metamaterials, plasmonic metamaterials, acoustic metamaterials, magnetic metamaterials, elastic metamaterials, terahertz metamaterials, and quantum metamaterials \cite{3dmeta}. They offer a significant platform for exploring analogous phenomena in high-energy physics, gravitation, astrophysics, and more. However, our work has demonstrated that the mapping between these branches and the metamaterials must be approached with caution, as this mapping must provide an adequate description of the problem.

In summary, our research provides valuable insights into the interplay between quantum field theory, spacetime geometry, and metamaterials, potentially paving the way for exciting new experimental directions in the study of thermal effects in various analogous physical systems.

\section*{Declaration of Competing Interest}

The authors declare no conflict of interest.

\section*{Acknowledgement}

This work was partially supported
by the Brazilian agencies CNPq and FAPEMIG. C. Filgueiras and L.C.T.Brito acknowledges FAPEMIG Grant No. APQ 02226/22. C. Filgueiras acknowledges CNPq Grant No. 310723/2021-3. R. M. Santos acknowledges FAPEMIG Grant No. 13681/2021-3.

\section*{Data Availability Statement}
No Data associated in the manuscript

\end{document}